\documentclass[useAMS,usenatbib]{mn2e}
\usepackage{graphicx,txfonts,amssymb,amsfonts,natbib,times,amsbsy}
\usepackage[]{hyperref}

\begin{document}

\title[Non-Gaussian cosmic shear skeleton]
  {The effect of primordial non-Gaussianity on the skeleton of cosmic shear maps}

\author[C.\,Fedeli et al.]{C.\,
      Fedeli$^{1,2}$, F.\, Pace$^{3}$, L.\, Moscardini$^{2,4,5}$, M. Grossi$^{6}$, and K. Dolag$^{6}$\\
      $^1$ Department of Astronomy, University of Florida, 312 Bryant Space Science Center, Gainesville, FL 32611 (cosimo.fedeli@astro.ufl.edu)\\
      $^2$ Dipartimento di Astronomia, Universit\`a di Bologna, Via Ranzani 1, I-40127 Bologna, Italy\\
      $^3$ Zentrum f\"ur Astronomie der Universit\"at Heidelberg, Institut f\"ur Theoretische Astrophysik, Albert \"Uberle Str. 2, D-69120 Heidelberg, Germany\\
      $^4$ INFN, Sezione di Bologna, Viale Berti Pichat 6/2, I-40127 Bologna, Italy\\
      $^5$ INAF - Osservatorio Astronomico di Bologna, Via Ranzani 1, I-40127 Bologna, Italy\\
      $^6$ Max-Planck-Institut f\"ur Astrophysik, Karl-Schwarzschild Stra\ss e 1, D-85748 Garching, Germany}
      
\maketitle

\begin{abstract}
We explore the imprints of deviations from Gaussian primordial density fluctuations on the skeleton of the large-scale matter distribution as mapped through cosmological weak lensing. The skeleton of the Large Scale Structure traces its filamentary structure and thus provides a complementary piece of information with respect to, e.g., the simple Probability Distribution Function or the peak number counts. We computed the skeleton length of simulated effective convergence maps covering $\sim 35$ sq. deg each, extracted from a suite of cosmological $n-$body runs with different levels of local primordial non-Gaussianity. The latter is expected to alter the structure formation process with respect to the fiducial Gaussian scenario, and thus to leave a signature on the cosmic web. We found that alterations of the initial conditions consistently modify both the cumulative and the differential skeleton length, although the effect is generically smaller than the cosmic variance and depends on the smoothing of the map prior to the skeleton computation. Nevertheless, the qualitative shape of these deviations is rather similar to their primordial counterparts, implying that skeleton statistics retain good memory of the initial conditions. We performed a statistical analysis in order to find out at what Confidence Level primordial non-Gaussianity could be constrained by the skeleton test on cosmic shear maps of the size we adopted. At $68.3\%$ Confidence Level we found an error on the measured level of primordial non-Gaussianity of $\Delta f_\mathrm{NL}\sim 300$, while at $90\%$ Confidence Level it is of $\Delta f_\mathrm{NL}\sim 500$. In both cases we ignored the effects of observational noise and degeneracy with other cosmological parameters. While these values by themselves are not competitive with the current constraints, weak lensing maps larger than those used here would have a smaller field-to-field variance, and thus would likely lead to tighter constraints. A rough estimate indicates $\Delta f_\mathrm{NL} \sim$ a few tens at $68.3\%$ Confidence Level for an all-sky weak lensing survey.\\
\\
\noindent
{\bf Key words:} cosmology: theory $-$ gravitational lensing: weak $-$ cosmological parameters $-$ large-scale structure of the Universe
\end{abstract}

\section{Introduction}\label{sct:introduction}

One of the most striking successes of the standard cosmological paradigm consists in explaining the structure formation process in the Universe. Amplification of quantum fluctuations during the early (inflationary) phase of expansion led to seed density perturbations in the dark matter. Overdense regions subsequently grew up due to gravitational instability, and eventually collapsed, detaching from the overall expansion of the Universe and forming the bound dark matter halos that today host galaxies and galaxy clusters. The standard model of inflation \citep{ST79.1,GU81.1,LI82.1} predicts a distribution of primordial density fluctuations that is indistinguishable from a Gaussian. This, coupled with the results of perturbation theory and cosmological numerical simulations, gives us a relatively complete picture for the formation of the Large Scale Structure (LSS henceforth). 

Current cosmological experiments however still leave room for alternative scenarios. Particularly, extensions of the most standard model of inflation can produce substantial deviations from a Gaussian distribution of primordial density and potential fluctuations (see \citealt{BA04.1,CH10.1,DE10.1} for recent reviews). The amount and shape of these deviations depend critically on the kind of non-standard inflationary model that one has in mind. Primordial non-Gaussianity is expected to modify the low-redshift LSS in different ways, most notably by shifting the timing for the collapse of the largest dark matter halos and inducing a model-specific scale dependence on the large scale bias thereof. Turning the argument around, observations of the LSS are critically important in order to validate or discard inflationary models, and thus get a better grasp on the physics of the early Universe.

As a matter of fact, the problem of constraining deviations from primordial Gaussianity by means different than the Cosmic Microwave Background (CMB) intrinsic temperature anisotropies has recently attracted much attention in the literature, with efforts directed towards the abundance of non-linear structures (\citealt*{MA00.2}; \citealt{VE00.1}; \citealt*{MA04.1}; \citealt{GR07.1,GR09.1,MA10.2}), halo biasing (e.g., \citealt{DA08.1,MC08.1}; \citealt*{FE09.1}; \citealt{FE10.1}), galaxy bispectrum (see, for instance, \citealt{SE07.2,JE09.1}), mass density distribution \citep{GR08.2}, cosmic shear (\citealt*{FE10.2}; \citealt{PA11.1}; \citealt*{MA11.1}), integrated Sachs-Wolfe effect \citep*{AF08.1,CA08.1}, Ly$\alpha$ flux from low-density intergalactic medium \citep{VI09.1}, $21-$cm fluctuations \citep*{CO06.2,PI07.1} and reionization \citep{CR09.1}.

The gravitational deflection of light is one of the most powerful tools to probe the LSS of the Universe. Gravitational lensing reacts only to the total gravitational potential integrated along the line of sight, thus it is insensitive to the intrinsic nature and dynamical status of the matter providing the deflection. In cosmological context it can be used to probe the large-scale matter distribution without relying on strong assumptions, as instead is the case for biased tracers such as galaxies and galaxy clusters. In the present paper we were interested in evaluating the effects of primordial non-Gaussianity on the topology of the (projected) matter distribution measured through cosmological weak lensing (cosmic shear).

Topological statistics such as the Minkowski Functionals of the LSS (\citealt{MA03.2,HI08.2}) and of the CMB maps \citep*{HI06.1} were already recognized as useful tools for constraining the cosmological initial conditions (see also \citealt*{MA01.2,VI09.2} and the early results of \citealt{ME90.1} and \citealt{MO91.1}). In this work we dealt with one particular morphological indicator of cosmic shear maps, namely the skeleton of the effective convergence. The skeleton is a set of connected lines that trace the filamentary structure of a given field, and the rigorous theory for the skeleton of a two and three-dimensional random field can be found in \citet{PO09.1} (see also \citealt*{NO06.1}; \citealt{PI10.1}). Models with a positive level of primordial non-Gaussianity are expected to provide an anticipated structure formation, with the consequence that at early times the filamentary structure is presumed to be more pronounced with respect to models with a negative skewness. At late times however, the matter is also expected to flow more efficiently toward the knots of the cosmic web, thus leaving the filaments relatively emptier in the former models compared to the latter. Understanding the net effect on the filamentary structure of cosmic shear maps is an interesting issue by itself.

The remainder of this work is structured as follows. In Section \ref{sct:ng} we describe the basics of cosmologies with non-Gaussian initial conditions and the $n-$body/ray-tracing simulations that have been used for the subsequent analysis. In Section \ref{sct:skeleton} we summarize how the skeleton of a two-dimensional field can be computed, and the relative local approximation that we used. In Section \ref{sct:results} we describe our results on the skeleton length in non-Gaussian cosmologies, and in Section \ref{sct:conclusions} we summarize our conclusions.

\section{Primordial non-Gaussianity}\label{sct:ng}

\subsection{Cosmological models}\label{sct:cosmology}

A particularly convenient (although not unique) way to describe generic deviations from Gaussian initial conditions consists in writing the gauge-invariant Bardeen's potential $\Phi$ as the sum of a Gaussian random field and a quadratic correction \citep{SA90.1,GA94.1,VE00.1,KO01.1}, according to

\begin{equation}\label{eqn:ng}
\Phi = \Phi_\mathrm{G} + f_\mathrm{NL} * \left( \Phi_\mathrm{G}^2 - \langle \Phi_\mathrm{G}^2 \rangle \right).
\end{equation}
The parameter $f_\mathrm{NL}$ in Eq. (\ref{eqn:ng}) determines the amplitude of non-Gaussianity, and it is in general dependent on the scale. The symbol $*$ denotes convolution between functions, and reduces to standard multiplication upon constancy of $f_\mathrm{NL}$. In the following we adopted the large-scale structure convention (as opposed to the CMB convention, see \citealt{AF08.1,CA08.1}; \citealt*{PI09.1} and \citealt{GR09.1}) for defining the fundamental parameter $f_\mathrm{NL}$. According to this, the primordial value of $\Phi$ has to be linearly extrapolated at $z = 0$, and as a consequence the constraints given on $f_\mathrm{NL}$ by the CMB have to be raised by $\sim 30$ per cent to comply with this paper's convention (see also \citealt*{FE09.1} for a concise explanation).

In the case in which $f_\mathrm{NL} \ne 0$ the potential $\Phi$ is a random field with a non-Gaussian probability distribution. Therefore, the field itself cannot be described by the power spectrum $P_\Phi({\boldsymbol k}) = Bk^{n-4}$ alone, rather higher order moments are needed. In many circumstances the dominant higher order moment is the bispectrum $B_\Phi(\boldsymbol{k}_1,\boldsymbol{k}_2,\boldsymbol{k}_3)$, that is the Fourier transform of the three-point correlation function $\langle \Phi(\boldsymbol{k}_1)\Phi(\boldsymbol{k}_2)\Phi(\boldsymbol{k}_3) \rangle$. It can be implicitly defined as

\begin{equation}
\langle \Phi(\boldsymbol{k}_1)\Phi(\boldsymbol{k}_2)\Phi(\boldsymbol{k}_3) \rangle \equiv (2\pi)^3\delta_\mathrm{D}\left( \boldsymbol{k}_1+\boldsymbol{k}_2+\boldsymbol{k}_3 \right) B_\Phi(\boldsymbol{k}_1,\boldsymbol{k}_2,\boldsymbol{k}_3).
\end{equation}

As mentioned above understanding the shape of non-Gaussianity is of fundamental importance in order to pinpoint the physics of the early Universe and the evolution of the inflaton field in particular. The standard single-field inflationary scenario generates negligibly small deviations from Gaussianity. These deviations are said to be of the local shape, and the related bispectrum of the Bardeen's potential is maximized for \emph{squeezed} configurations, where one of the three wavevectors has much smaller magnitude than the other two. In this case the parameter $f_\mathrm{NL}$ must be a constant, and it is expected to be of the same order of the slow-roll parameters \citep*{FA93.1}, that are very close to zero. 

However non-Gaussianities of the local shape can also be generated in the case in which an additional light scalar field, different from the inflaton, contributes to the observed curvature perturbations \citep*{BA04.2}. This happens, for instance, in curvaton models (\citealt*{SA06.1}; \citealt*{AS07.1}) or in multi-fields models \citep*{BA02.2,BE02.1}. In this case the parameter $f_\mathrm{NL}$ is allowed to be substantially different from zero, and the bispectrum of the primordial potential assumes the simple form

\begin{equation}
B_\Phi(\boldsymbol{k}_1,\boldsymbol{k}_2,\boldsymbol{k}_3) = 2f_\mathrm{NL} B^2 \left[ k_1^{n-4}k_2^{n-4} + k_1^{n-4}k_3^{n-4} + k_2^{n-4}k_3^{n-4} \right].
\end{equation}

As naively expected, the mass function of dark matter halos gets modified in non-Gaussian cosmologies. Several prescriptions have been developed over the years in order to describe this modification (\citealt*{MA00.2}; \citealt{LO08.1}; \citealt{DA11.1}), and all of them are in broad agreement with numerical simulations of structure formation (\citealt{GR09.1}; \citealt*{WA10.1}). Only quite recently was it recognized that primordial non-Gaussianity introduces a scale dependence on the linear bias of dark matter halos (\citealt{DA08.1}; \citealt*{MA08.1}). Such a dependence is regulated by suitable integrals of the potential bispectrum, and is one of the most promising LSS probes of primordial non-Gaussianity \citep*{AF08.1,CA08.1,VE09.1,CA10.1}.

\subsection{Numerical simulations}\label{sct:simulations}

The $n-$body simulations with non-Gaussian initial conditions that we employed are originally described in \citet{GR07.1,HI08.2,GR08.2}. We thus refer to those papers for detailed information, limiting ourselves to describe here only the main features. Dark matter-only structure formation was simulated by using the publicly available {\sc Gadget 2} code \citep{SP05.1}, assuming a flat $\Lambda$CDM cosmological model with density parameters for matter, baryons and cosmological constant equal to $\Omega_{\mathrm{m},0}=0.30$, $\Omega_{\mathrm{b},0}=0.04$, and $\Omega_{\Lambda,0}=0.70$, respectively. The Hubble constant was set to $H_0 = h 100$ km s$^{-1}$ Mpc$^{-1}$ with $h=0.7$, while the matter power spectrum was normalized according to the constraint $\sigma_8 = 0.9$. The comoving side length of the simulation box measures $500 h^{-1}$ Mpc, and the number of particles used is $800^3$, implying a mass resolution of $2\times 10^{10} h^{-1}M_\odot$. In addition to the Gaussian case $f_\mathrm{NL}=0$, six additional non-Gaussian scenarios were simulated, with the local shape for the primordial bispectrum described in the previous subsection and $f_\mathrm{NL} = \pm100, \pm 500$, and $\pm 1000$. It should be noted that current constraints on $f_\mathrm{NL}$ arising from the CMB and the LSS for the local case are $\Delta f_\mathrm{NL} \lesssim 100$, thus only the non-Gaussian simulations with $f_\mathrm{NL} = \pm 100$ can be considered realistic. Nevertheless, we performed our analysis for all the $f_\mathrm{NL}$ values at our disposal both for illustrative purposes and because this allows us to better interpolate the results for other values.

The ray-tracing analysis of the simulation cubes and the subsequent production of light cones and cosmic shear maps is described in detail in \citet{PA11.1}. Again, we merely summarize the procedure here. As a first step, the particles in the simulation boxes at each redshift snapshot have been randomly shifted and rotated in order to avoid repetitions of the same structure at different redshifts. Then, the particles themselves have then been projected along the line of sight on a plane lying in the middle of the box in order to compute the surface mass density on a regular grid. By the use of the two-dimensional Poisson equation this surface mass density has been subsequently converted into the lensing potential for the considered lens plane, from which it is possible to piecewise reconstruct the trajectory of each light ray shot from the observer toward the sources. Here all sources have been placed at the same redshift, $z_\mathrm{s} = 4$. Although this might be not entirely realistic, we are here just interested in the relative differences between Gaussian and non-Gaussian cosmologies. The relation between the true and apparent positions of sources allows to compute a variety of lensing observables, including shear, flexion, and effective convergence.

In this work we focused on the effective convergence only for two reasons: $i)$ it is more directly related to the projected matter distribution than, e.g., the shear, and thus it should have a more direct link to the initial conditions; $ii)$ as we verified, the corresponding skeleton tends to be more regular than the skeleton of other weak lensing observables. The (two-dimensional) effective convergence can indeed be related to the three-dimensional matter distribution as

\begin{equation}
\kappa_\mathrm{e}(\boldsymbol{\theta}) = \frac{3H_0^2\Omega_{\mathrm{m},0}}{2c^2}\int_0^{w_\mathrm{s}} dw \frac{f_K(w)f_K(w_\mathrm{s}-w)}{f_K(w_\mathrm{s})}\frac{\delta\left[f_K(w)\boldsymbol{\theta},w\right]}{a(w)},
\end{equation}
where $c$ is the speed of light, $w=w(z)$ is the comoving distance out to redshift $z$, $a(w)$ is the corresponding scale factor, $\delta(\boldsymbol{x})$ is the (dark) matter density contrast at a given spatial position $\boldsymbol{x}$, and $f_K(w_\mathrm{s}-w)$ is the comoving angular diameter distance between the redshift $z$ corresponding to $w(z)$ and the source redshift, depending on the spatial curvature $K$ (which is zero in our case).

From each of the seven $n-$body cosmological simulations described above we extracted $60$ effective convergence maps, each one obtained with a different randomization of the particle positions in the simulation boxes. They can thus be regarded as $60$ independent realizations of the same light cone, allowing a mean to estimate the effect of cosmic variance. Each map has a resolution of $2048^2$ gridpoints, and has an opening angle of $\sim 5.9$ deg, corresponding to a covered area of $\sim 35$ sq. deg.

\section{The skeleton of a two-dimensional field}\label{sct:skeleton}

The skeleton of a two-dimensional field $\rho$ is visually intended as the set of lines that trace the filamentary structure of the field itself. This naive definition can be made rigorous \citep*{NO06.1} by saying that the skeleton lines are parametrized by those trajectories ${\bf x}(t) \equiv  \left[x_1(t),x_2(t)\right]$ on the plane that obey the equation of motion

\begin{equation}
\frac{d{\bf x}}{dt} = \nabla \rho,
\end{equation}
starting at a saddle point and with initial velocity parallel to the direction of minimum curvature. This definition is non-local, and therefore difficult to be implemented practically. Fortunately, an acceptable local approximation of the skeleton can be devised \citep{PO09.1,PI10.1}. This local skeleton can be defined as a subset of the critical lines of the field, i.e. lines where the gradient of the field $\rho$ is parallel to one of the eigenvectors of the Hessian matrix. If $\lambda$ is an arbitrary eigenvalue of the Hessian $\mathcal{H}_\rho$, then this condition translates into

\begin{equation}\label{eqn:local}
\mathcal{H}_\rho\nabla\rho = \lambda\nabla\rho.
\end{equation}

By adopting the index notation, Eq. (\ref{eqn:local}) can be rewritten in a more useful form as

\begin{equation}\label{eqn:index}
\rho_{ij}\rho_j = \lambda\rho_i,
\end{equation}
where $\rho_i = \partial \rho/\partial x_i$, $\rho_{ij} = \partial^2\rho/\partial x_i \partial x_j$, and summation over repeated indices is implicit. The two conditions in Eq. (\ref{eqn:index}) can be rewritten in a more insightful way, as

\begin{equation}\label{eqn:c1}
\rho_{11} + \rho_{12}\frac{\rho_2}{\rho_1} = \lambda,
\end{equation}

\begin{equation}\label{eqn:c2}
\rho_{22} + \rho_{12}\frac{\rho_1}{\rho_2} = \lambda.
\end{equation}
By subtracting Eq. (\ref{eqn:c2}) from Eq. (\ref{eqn:c1}) we obtain after some manipulation that

\begin{equation}\label{eqn:c3}
S\equiv\rho_1\rho_2\left(\rho_{11}^{\phantom{1}}-\rho_{22}\right) + \rho_{12}\left( \rho_2^2-\rho_1^2 \right) = 0.
\end{equation}

Note that Eq. (\ref{eqn:c3}) is valid whenever the gradient of the field $\rho$ is parallel to one eigenvector of the Hessian, irrespective of which eigenvalue we are referring to. The relevant eigenvalue is specified only in the pair of Equations (\ref{eqn:c1}) and (\ref{eqn:c2}). The same Eq. (\ref{eqn:c3}) can also be obtained along a different route, by simply imposing the condition that the gradient of the field is parallel to an arbitrary eigenvalue of the Hessian, i.e.,

\begin{equation}
\mathcal{H}_\rho\nabla\rho \times \nabla\rho \equiv \varepsilon_{ij} \left( \mathcal{H}_{\rho,ik} \rho_k\right) \rho_j = 0,
\end{equation}
where $\varepsilon_{ij}$ is the fully antisymmetric two-dimensional Levi-Civita tensor.

The local skeleton can now be defined as the subset of the critical lines for which $\lambda$ is the largest eigenvalues, i.e. the eigenvalue related to the direction of minimal curvature at a given point, and the smallest eigenvalue is negative. If we label the two eigenvalues of the Hessian matrix as $\lambda_1$ and $\lambda_2$, and assume without loss of generality that $\lambda_1\ge \lambda_2$, then the local skeleton is defined by Eqs. (\ref{eqn:c1}) and (\ref{eqn:c2}) by setting $\lambda=\lambda_1$, with the supplemental condition that $\lambda_1+\lambda_2 \le 0$. As can be easily seen, since $\lambda_1\ge\lambda_2$, the latter condition implies that $\lambda_2\le 0$, irrespective of $\lambda_1$.

\begin{figure}
\begin{center}
	\includegraphics[width=0.85\hsize]{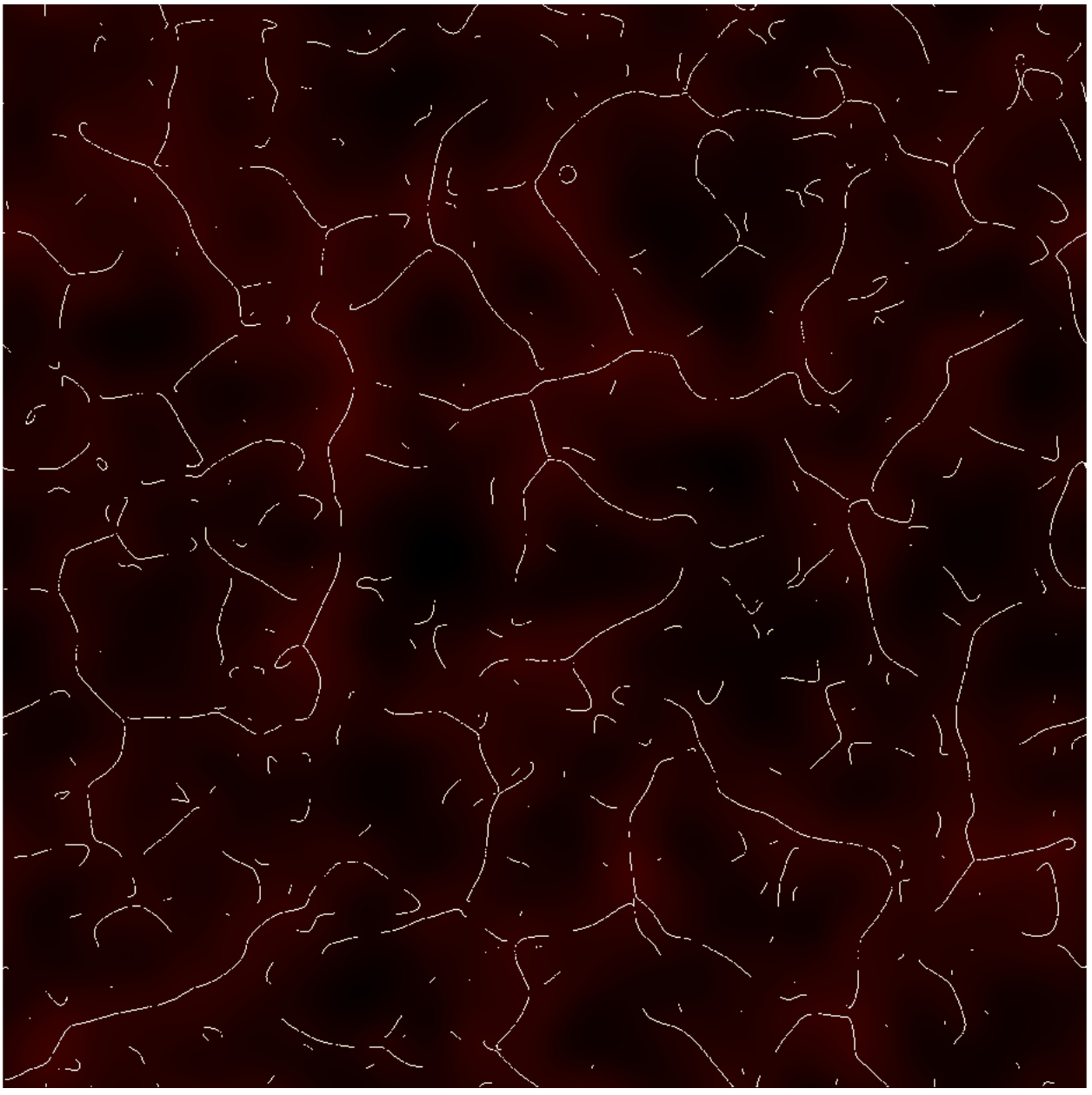}\hfill
	\caption{The effective convergence map of a randomly chosen light cone realization for the Gaussian simulation ($f_\mathrm{NL}=0$), smoothed with a Gaussian kernel whose \emph{rms} is $\ell = 0.02$ times the side length of the map. The overimposed white lines represent the local skeleton of the smoothed field, as defined in the text. The side length measures $\sim 5.9$ deg.}
\label{fig:skeleton}
\end{center}
\end{figure}

In order to ensure the differentiability of the field under consideration up to second order, it is custumary to smooth the field itself prior to the calculation of the local skeleton. It is immediate to understand that the length of the skeleton will depend on the smoothing length, with smoother fields having less features, and thus a shorter skeleton, with respect to less smooth ones. In this work we adopted a Gaussian smoothing kernel, and performed calculations for two different smoothing scales: $\ell = 0.02$ and $\ell=0.002$ times the side length of the maps. These values correspond to $\ell \sim 7$ arcmin and $\ell\sim 43$ arcsec, respectively. In the remainder of this work we will show alternatively the results obtained with the first or with the second smoothing scale, depending on which one fits best the purpose of illustration.

As a demonstration, in Figure \ref{fig:skeleton} we show the local skeleton of the effective convergence for one of the $60$ light cone realizations in the reference Gaussian cosmology. The map has been smoothed with a Gaussian kernel whose \emph{rms} is $\ell=0.02$ times the side length of the box. We observe that, since the smoothing has been performed with a Fast Fourier Transform (FFT) algorithm, the resulting smoothed map has periodic boundaries. We also note two additional facts: \emph{i)} the local skeleton follows quite nicely the ridges of the effective convergence map, highlighting the filamentary structure thereof; \emph{ii)} the local skeleton itself is quite irregular, and often broken in many pieces. This is a consequence of the local approximation, since the true skeleton would be made of continuos lines \citep*{NO06.1}.

The main statistics we will be interested in for the remainder of this paper is the length of the local skeleton related to excursion maps. Excursion maps are subsets of the original map given by considering only those points where the field value exceed some threshold value, $\rho \ge \rho_\mathrm{t}$. For subsequent use, we defined a new field $\eta$ that is given by the original field divided by its \emph{rms}, that is $\eta = \rho/\sigma$, where $\sigma = \sqrt{\langle \rho^2 \rangle-\langle \rho \rangle^2}$. Under this rescaling the threshold field value maps into a threshold $\eta$ value, $\eta_\mathrm{t} = \rho_\mathrm{t}/\sigma$. As in Figure \ref{fig:skeleton} and in the remainder of this work, $\rho = \kappa_\mathrm{e}$, where $\kappa_\mathrm{e}$ is the effective lensing convergence. We stress that, since the effective convergence is an integral along the line of sight of the density contrast, it may also take negative values.

\begin{figure}
\begin{center}
	\includegraphics[width=\hsize]{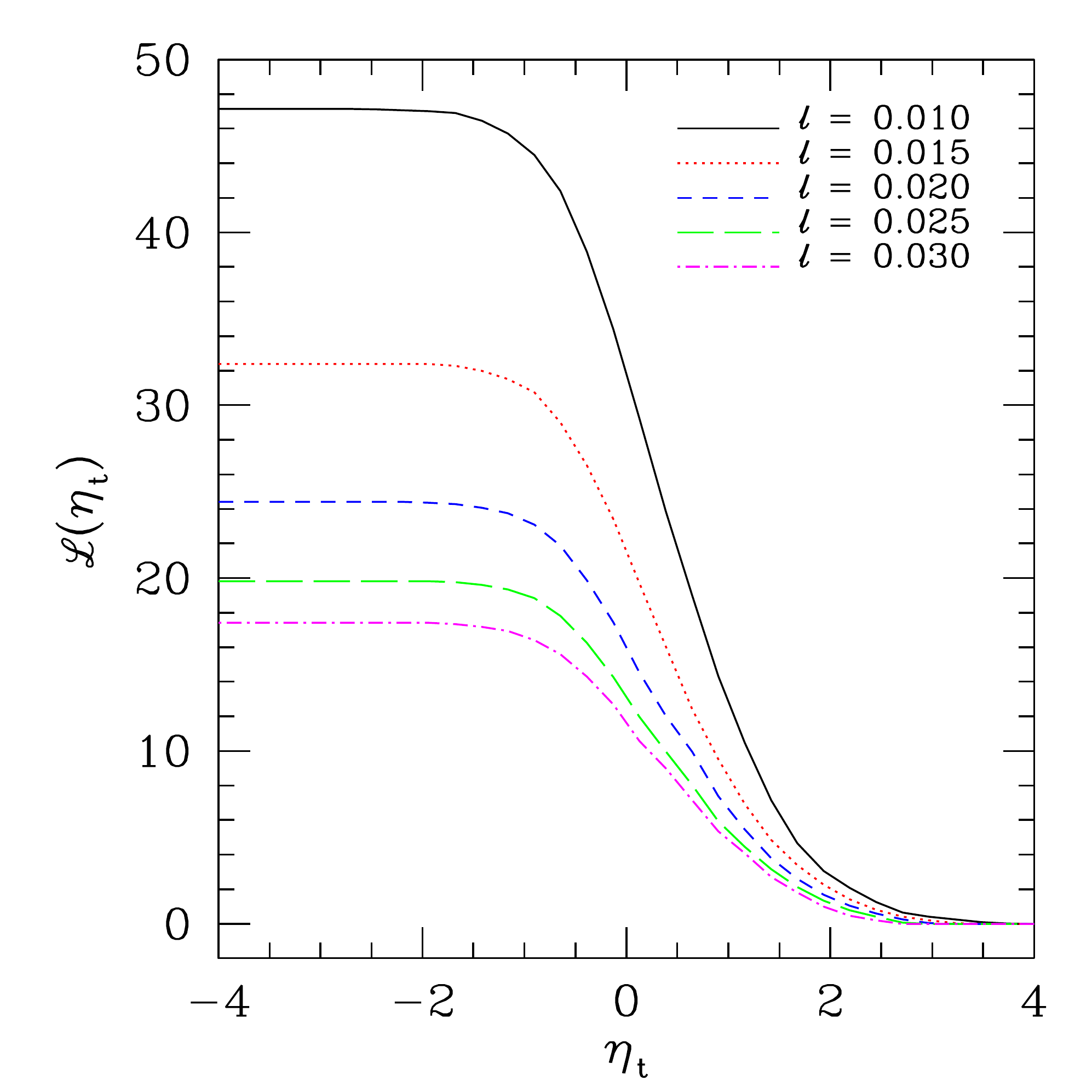}\hfill
	\caption{The skeleton length for excursion maps constructed from the effective convergence map shown in Figure \ref{fig:skeleton} ($f_\mathrm{NL}=0$), as a function of the threshold value in units of the field \emph{rms}. The skeleton length is expressed in units of the side length of the map ($\sim 5.9$ deg), and five different sizes of the Gaussian smoothing kernel are adopted, as labeled in the plot.}
\label{fig:test}
\end{center}
\end{figure}

\begin{figure*}
\begin{center}
	\includegraphics[width=0.45\hsize]{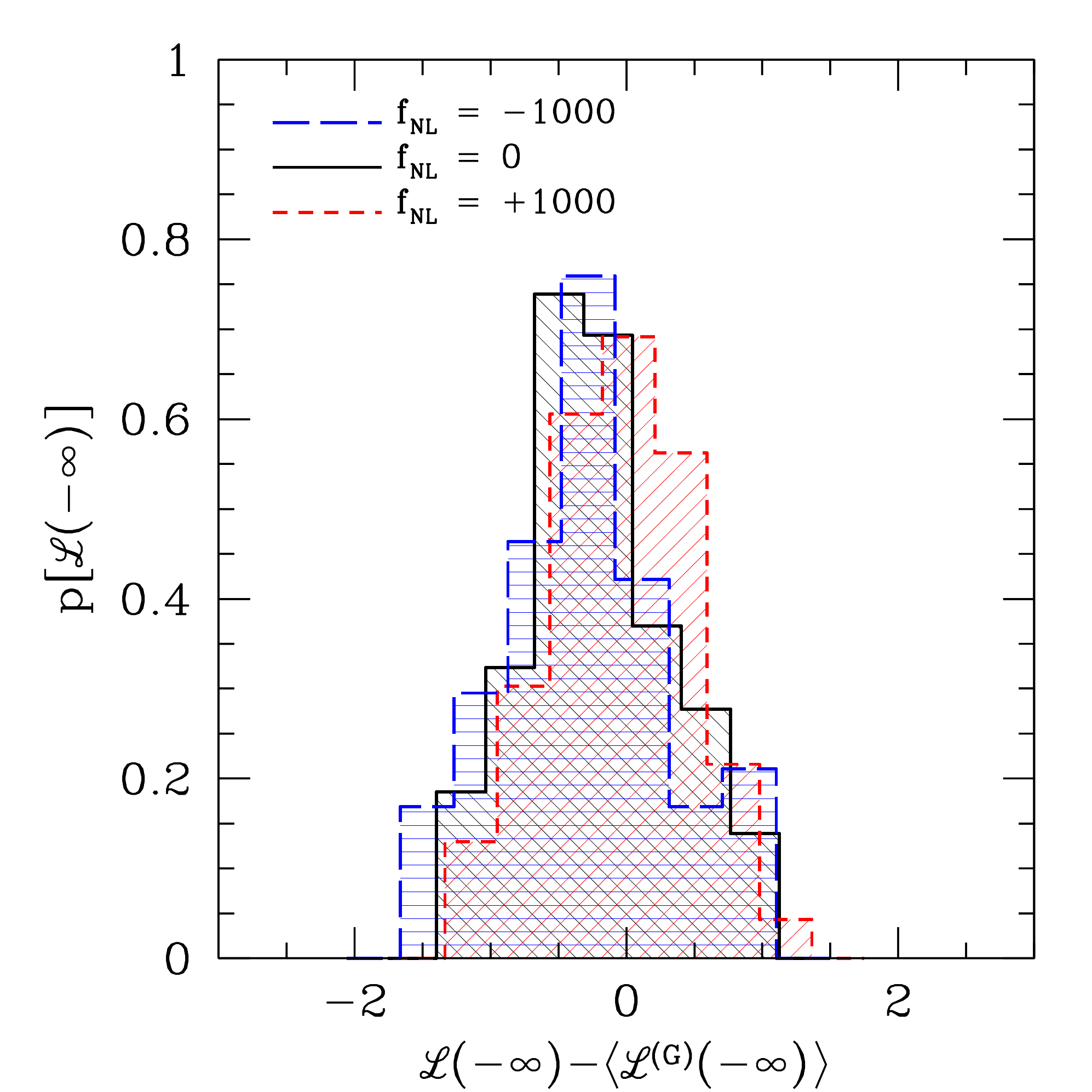}
	\includegraphics[width=0.45\hsize]{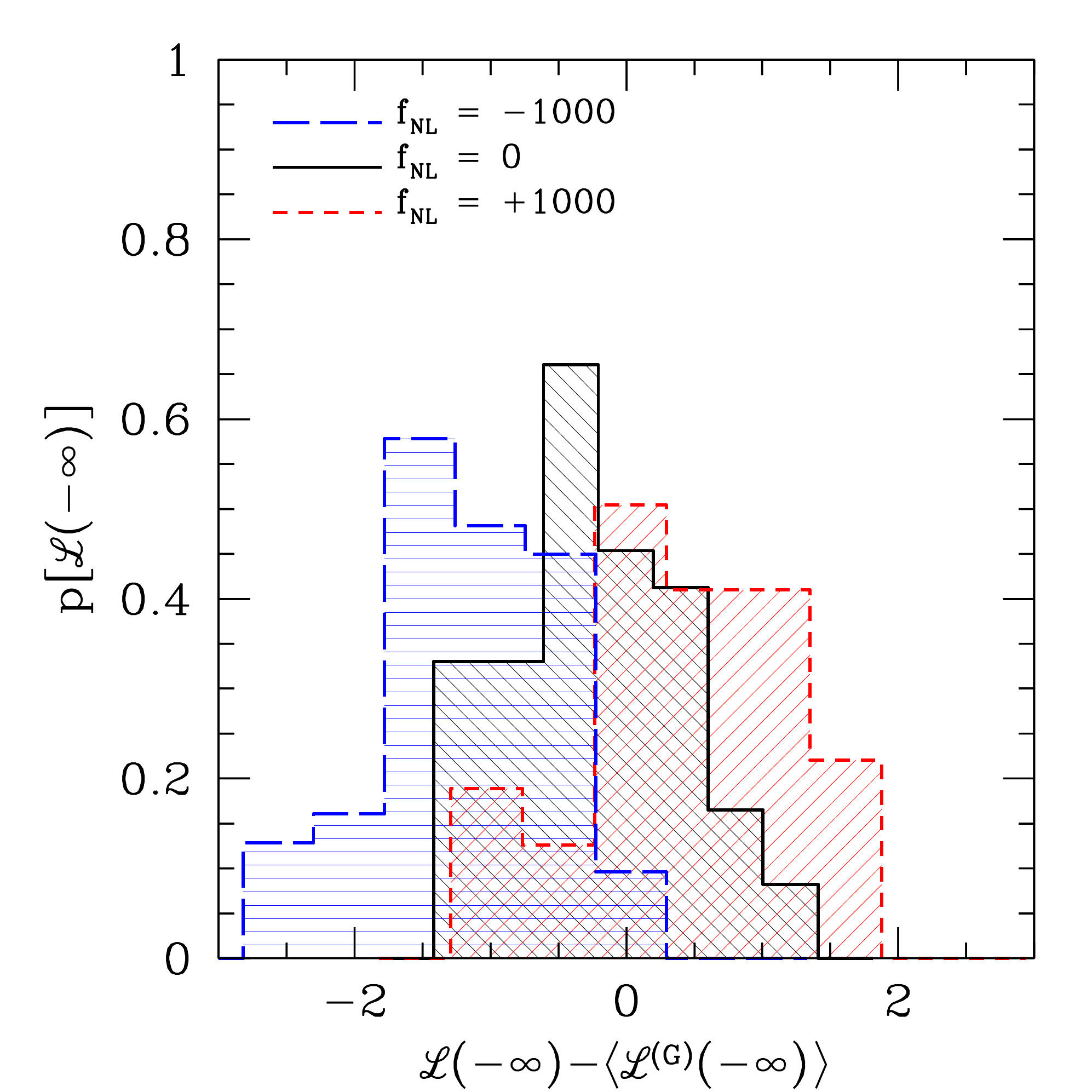}\hfill
	\caption{The distribution of the total skeleton length (in units of the map side length) for effective convergence maps smoothed with a Gaussian kernel of width $\ell=0.02$ times the side length ($\sim 7$ arcmin, left panel) and $\ell=0.002$ times the side length ($\sim 43$ arcsec, right panel). Distributions are computed over the $60$ light cone realizations, and results for three different values of the non-Gaussian parameter $f_\mathrm{NL}$ are shown, as labeled in the plots. Note that the abscissa in each panel is shifted according to the average value of the total skeleton length in the Gaussian case.}
\label{fig:absoluteLengthDistribution}
\end{center}
\end{figure*}

Measuring the length of the skeleton is an issue that deserves some further discussion. In other works (\citealt{ER04.1}; \citealt*{NO06.1}; \citealt{HO10.1}) this length has been measured by using linear or cubic interpolation in order to find where a skeleton line effectively crosses the borders of a pixel that is known to host a critical curve. Then, it was assumed that the curve itself is a line segment between the crossing points on two sides of the pixel. Of course, this reasoning does not work when multiple critical lines converge to the same point (critical point) within one pixel or when the critical curves are interrupted, an instance that is often verified for the local approximation of the skeleton at our resolution level (see Figure \ref{fig:skeleton}). Thus, since some level of approximation is unavoidable, we chose to adopt a simplified approach, and assign to each pixel a length scale that equals its side length. The average length of a line segment randomly crossing a squared pixel is probably somehow different from the side length of the pixel. However, we are not interested in the absolute value of the skeleton length, rather on how this and its differential and cumulative counterparts are modified in presence of primordial non-Gaussianity. Thus, as long as we adopt the same convention for all maps in all simulations, our conclusions should hold.

In Figure \ref{fig:test} we show, for the same effective convergence realization represented in Figure \ref{fig:skeleton}, the length of the local skeleton for excursion maps as a function of the threshold $\eta_\mathrm{t}$ used to define the excursion map itself. We name this quantity as the cumulative skeleton length $\mathcal{L}(\eta_\mathrm{t})$, while the differential skeleton length is its derivative normalized by the length of the skeleton of the original map, $\mathcal{D}\left(\eta_\mathrm{t}\right)\equiv-\left[1/\mathcal{L}(-\infty)\right]d\mathcal{L}(\eta_\mathrm{t})/d\eta_\mathrm{t}$. Another quantity we shall be interested with is the normalized cumulative skeleton length, namely $\mathcal{C}(\eta_\mathrm{t}) \equiv \mathcal{L}(\eta_\mathrm{t})/\mathcal{L}(-\infty)$. In Figure \ref{fig:test} we report curves for several different values of the Gaussian smoothing scale around $\ell=0.02$. As one could naively expect, increasing the smoothing scale decreases the length of the skeleton, since the more the field is smooth, the less features it displays.

\section{Results}\label{sct:results}

\subsection{Total skeleton length}

We now turn attention on how the skeleton is modified by the introduction of some level of primordial non-Gaussianity. The first thing that we have investigated is how the total length of the local skeleton (without excursion) does change. It is expected that primordial non-Gaussianity with a positive skewness of the matter density field would produce an early enhancement of the filamentary structure in the large scale matter distribution, that would be later compensated by the more efficient growth of dark matter clumps. On the other hand, a negative skewness would produce the opposite effect.

In the left panel of Figure \ref{fig:absoluteLengthDistribution} we show the distribution of the total length of the local skeleton $\mathcal{L}(-\infty)$ obtained by the $60$ effective convergence maps we studied. In order to produce this panel all maps have been smoothed with a Gaussian kernel whose \emph{rms} is $\ell = 0.02$ times the side length of the map (corresponding to $\sim 7$ arcmin). Besides result for the fiducial Gaussian cosmology, we also report those for the most extreme non-Gaussian simulations, with $f_\mathrm{NL} = \pm 1000$. Despite having adopted the largest $f_\mathrm{NL}$ values at our disposal, it is apparent that very little difference is present between the three distributions, which overlap with each other almost completely. It is true that the average value of the distribution referring to $f_\mathrm{NL} = +1000$ is larger than that relative to $f_\mathrm{NL} = -1000$, implying an effective net enhancement of the filamentary structure, however the difference between the two is only $\sim 0.25$ (times the side length of the map), while the \emph{rms} of the individual distributions is $\sim 0.6$. 

\begin{figure*}
\begin{center}
	\includegraphics[width=0.45\hsize]{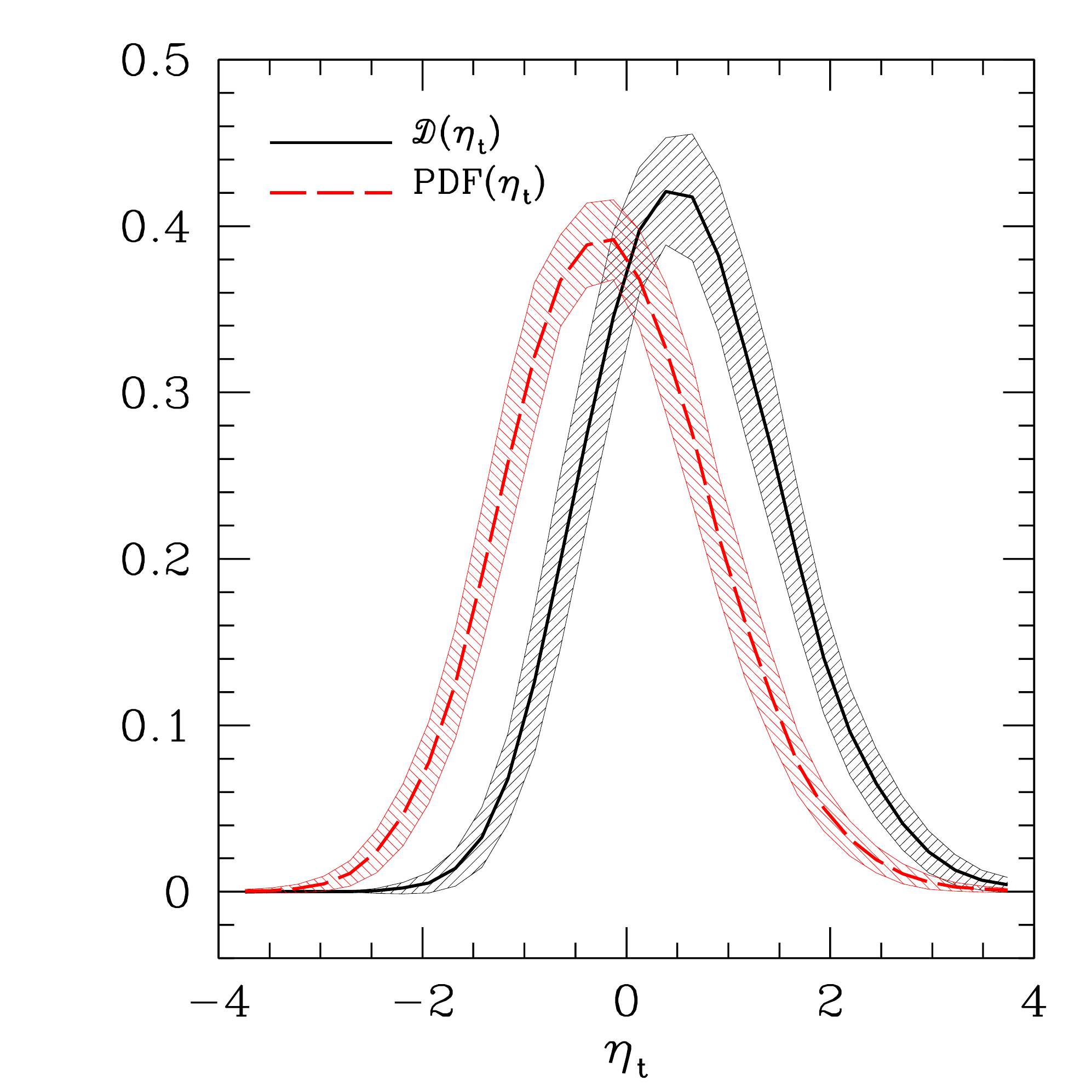}
	\includegraphics[width=0.45\hsize]{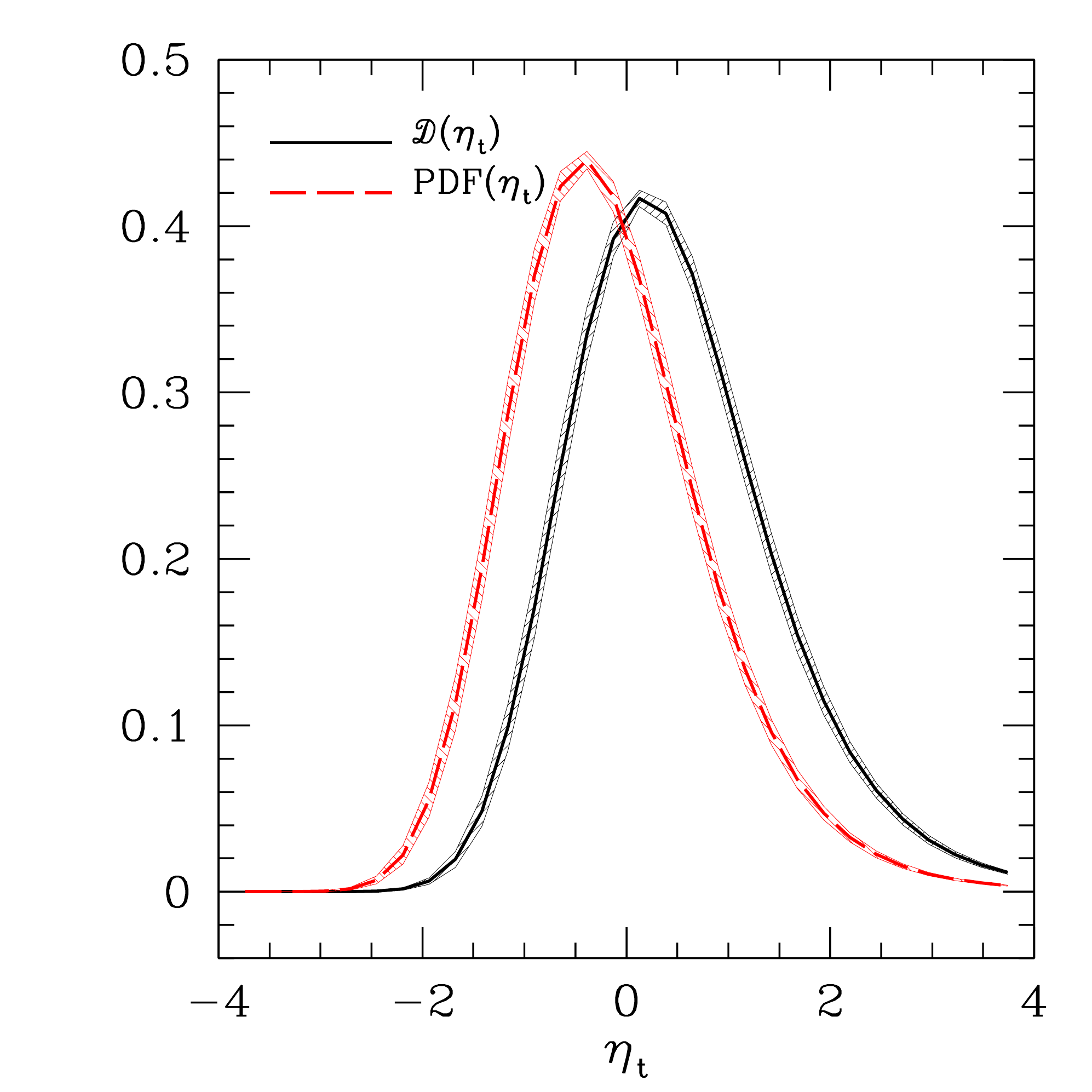}\hfill
	\caption{The differential skeleton length distribution (black solid line) and the PDF (red dashed line) of the effective convergence for the Gaussian cosmology. The lines show the average values computed over the $60$ light cone realizations as labeled, while the shaded regions represent the dispersions around the means of the distributions for each $\eta_\mathrm{t}$. Each effective convergence map has been smoothed with a Gaussian kernel of radius $\ell = 0.02$ (left panel) and $\ell = 0.002$ times its side length (right panel).}
\label{fig:differentialLengthDistribution}
\end{center}
\end{figure*}

In the right panel of Figure \ref{fig:absoluteLengthDistribution} we show the same results, but this time assuming a smaller smoothing length of $\ell=0.002$ times the side of the map (corresponding to $\sim 43$ arcsec), in order to check whether a larger difference in the total skeleton lengths could arise upon inclusion of smaller scale features of the effective convergence maps. As it turns out, this is indeed the case: the three distributions show now a substantial separation in the same direction noted above. In this case the average values of the two distributions referring to the cases $f_\mathrm{NL}=\pm 1000$ differ by $\sim 1.6$ times the side of the map, while the \emph{rms} is $\sim 0.8$ in the same units. This result is expected, and can be explained as follows. Primordial non-Gaussianity tend to affect more strongly the most extreme features of the LSS, for instance the high-density peaks and ridges. The effective convergence we are dealing with is actually a projection of the LSS, and the larger is the smoothing radius, the more these extreme features are washed out from the maps, thus effectively reducing the leverage of the skeleton to tell the models apart. A simple Kolmogorov-Smirnov test implies the two distributions for $f_\mathrm{NL}=-1000$ and $f_\mathrm{NL} = +1000$ being the same at more than $20\%$ Confidence Level for $\ell = 0.02$ and only at less than $0.5\%$ Confidence Level for $\ell = 0.002$, further confirming the result of the visual inspection.

The fact that the dispersions of the three distributions shown in Figure \ref{fig:absoluteLengthDistribution} is larger for $\ell = 0.002$ than for $\ell = 0.02$ does not mean that the total skeleton length is more affected by the cosmic variance in the former case, because $\left\langle \mathcal{L}^\mathrm{(G)}(-\infty) \right\rangle$ is much larger there than in the latter. As we will detail in the next Subsection, the normalized cumulative and differential skeleton lengths, that are correctly normalized, actually show the opposite behavior.

\subsection{Differential and cumulative skeleton lengths}

The next step is to deal with the differential length of the local skeleton $\mathcal{D}(\eta_\mathrm{t})$ and its cumulative counterpart $\mathcal{C}(\eta_\mathrm{t})$. We start with the former. As noted in \citet*{NO06.1} the differential skeleton length should behave quite similarly to the Probability Distribution Function (PDF henceforth) of the field, although several subtle differences exist that make it worth studying \citep{HO10.1}. One such difference is made by the fact that we are considering the skeleton of the effective convergence, but not the anti-skeleton (that would follow the valleys of the field instead of the ridges, and can be viewed as the skeleton of $-\kappa_\mathrm{e}$, see \citealt{PO09.1}). As a consequence the differential length distribution is expected to be more skewed toward high effective convergence values as compared to the simple PDF. This is an advantageous feature for our purposes, since primordial non-Gaussianity has indeed more effect on the more extreme details (e.g., ridges) of the LSS. Although not included here, the anti-skeleton by itself could share equally advantageous features, since primordial non-Gaussianity is expected to affect the most extreme voids as well. Even if the anti-skeleton was included in the analysis however, the critical lines would still sample the field in a non-random way, at least if we are not in the peculiar, completely incoherent case. Thus, differences between the PDF and the differential length are expected anyway \citep*{NO06.1}.

In Figure \ref{fig:differentialLengthDistribution} we report the differential length distribution of the skeleton $\mathcal{D}(\eta_\mathrm{t})$, comparing it with the PDF of effective convergence maps in the reference Gaussian cosmological model. As expected, the differential length of the local skeleton is more shifted to high values of the effective convergence with respect to the PDF. This is in agreement with the naive expectations that we outlined above. Particularly, the peak of the differential length is at $\sim 0.5\sigma$, while the peak of the PDF is only at $\sim -0.5\sigma$. In the same Figure we report results for the two different smoothing lengths adopted in this work, $\ell = 0.02$ and $\ell = 0.002$ times the side length of the maps. The behaviors of the distributions in the two cases are relatively similar. Specifically, the positions of the peaks are approximately unchanged. The only significant difference is given by the dispersion around the mean for a fixed $\eta_\mathrm{t}$, which is substantially larger in the case $\ell = 0.02$ than in the case $\ell = 0.002$. This behavior is expected because for smoother fields the typical length scale of features tend to be larger, thus they are presumed to be more subject to field-to-field variance.

\begin{figure}
\begin{center}
	\includegraphics[width=\hsize]{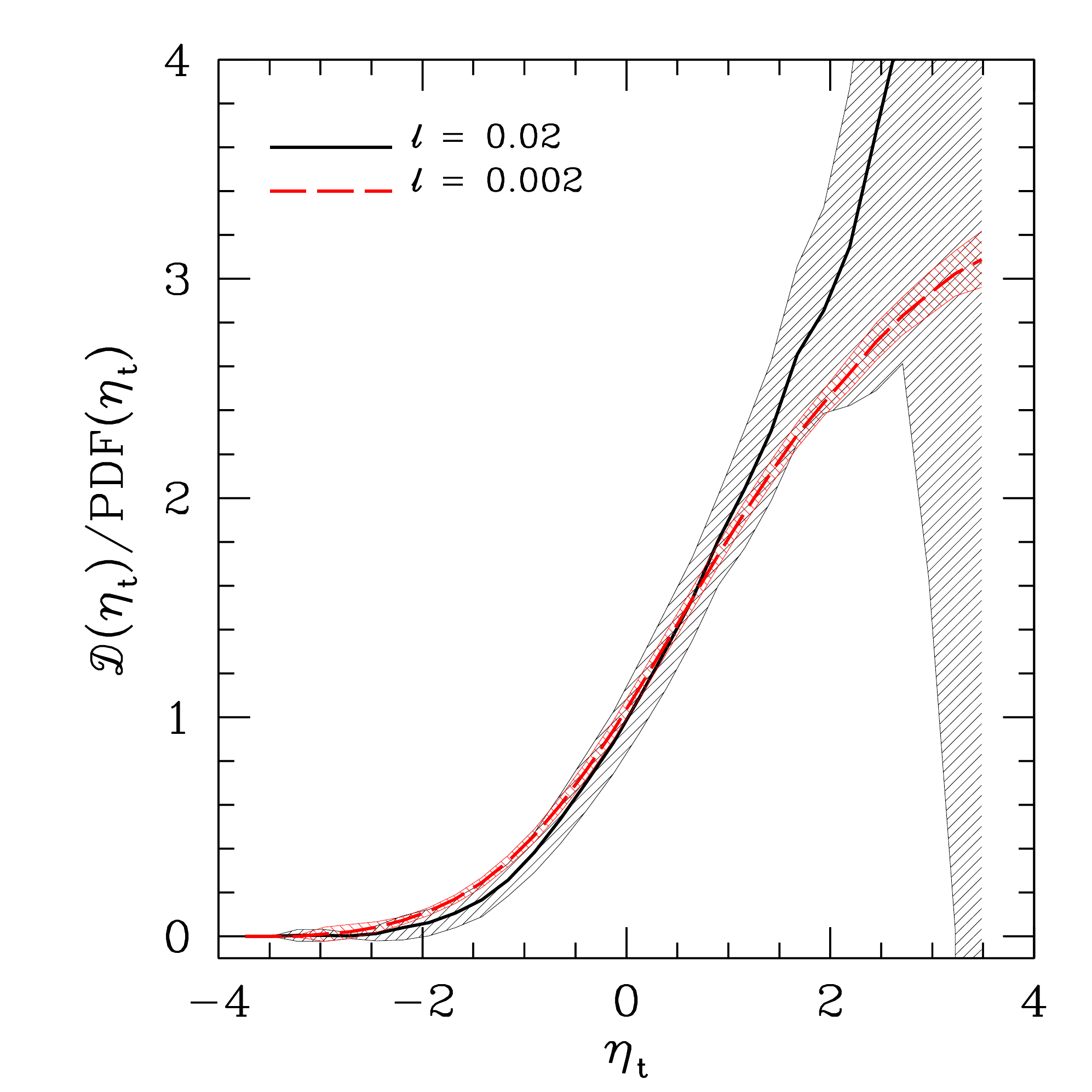}\hfill
	\caption{The distribution of the ratio of the differential skeleton length to the PDF of the effective convergence in the Gaussian simulation. For each value of $\eta_\mathrm{t}$ the average value over the $60$ light cone realizations is shown by the black solid line for $\ell = 0.02$ and by the red dashed line for $\ell = 0.002$, as labeled in the plot. The shaded regions show instead the dispersions around the mean values.}
\label{fig:differentialRatioDistribution}
\end{center}
\end{figure}

An alternative way to display the difference between the PDF of the effective convergence field and its differential skeleton length can be seen in Figure \ref{fig:differentialRatioDistribution}, where we show the distribution of the ratio of $\mathcal{D}(\eta_\mathrm{t})$ to PDF$(\eta_\mathrm{t})$ in the fiducial Gaussian cosmological model. As a consequence of the tendency of the local skeleton to trace the ridges of the map, this ratio increases steadily with increasing excursion threshold $\eta_\mathrm{t}$. The trend of the ratio is qualitatively the same that has been shown in \citet{PO09.1} for a two-dimensional Gaussian random field, although the field we are considering is not Gaussian, even if the initial conditions are, due to the non-linear clustering of dark matter.

\begin{figure}
\begin{center}
	\includegraphics[width=\hsize]{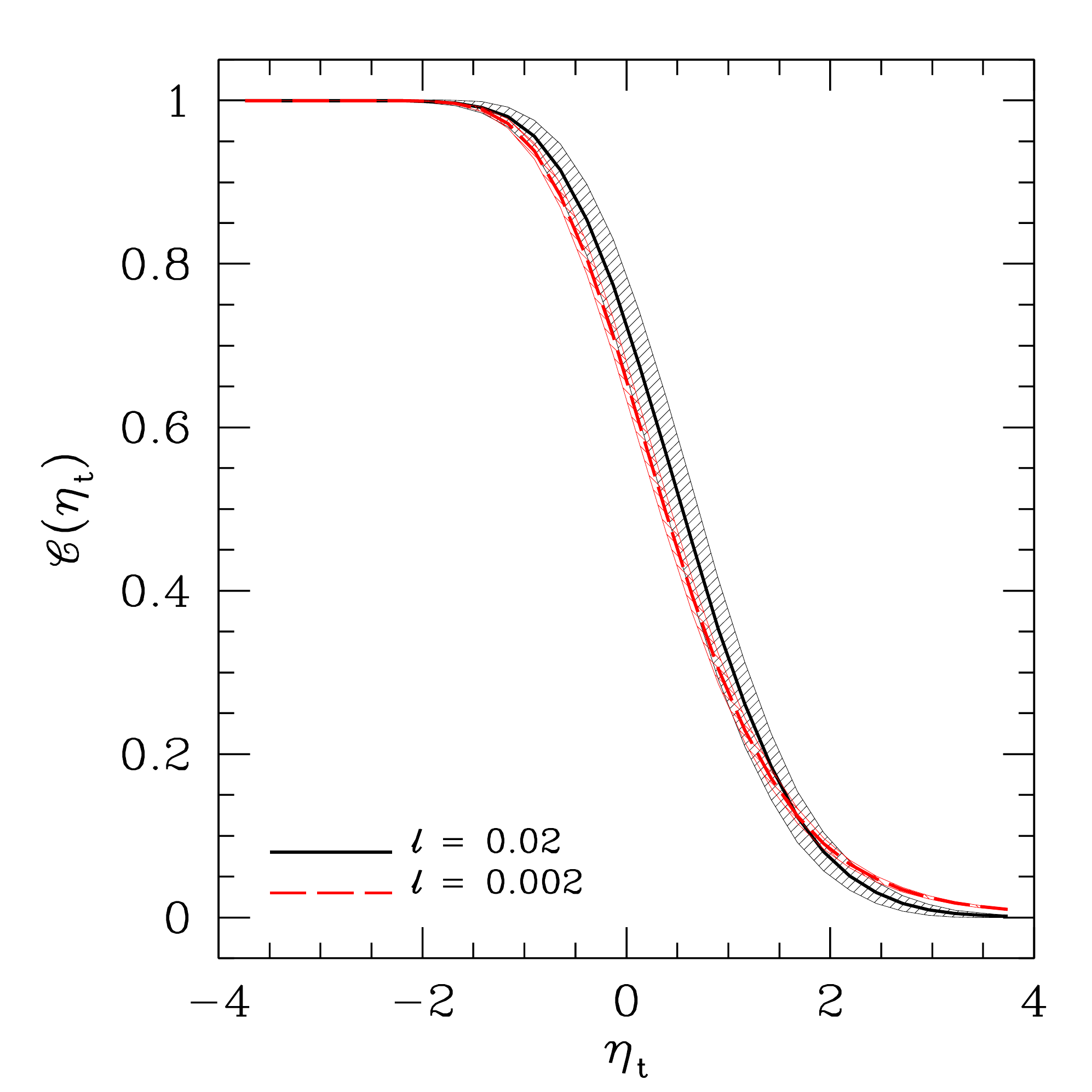}\hfill
	\caption{The normalized cumulative skeleton length distribution of the effective convergence for the Gaussian cosmology. The lines show the average value over the $60$ light cone realizations, while the shaded regions represent the dispersion around the mean of the distribution for each $\eta_\mathrm{t}$. Each effective convergence map has been smoothed with a Gaussian kernel of radius $\ell = 0.02$ and $\ell = 0.002$ times its side length, as labeled in the plot.}
\label{fig:cumulativeLengthDistribution}
\end{center}
\end{figure}

The normalized cumulative skeleton length distribution for the fiducial Gaussian cosmology is shown in Figure \ref{fig:cumulativeLengthDistribution} for the two Gaussian smoothing scales that we adopted in this work. The behavior is the one expected by looking at Figure \ref{fig:test}, with only little difference between the two smoothing radii adopted. Also in this case the only significant difference is given by the dispersion around the average value, that is substantially larger when $\ell = 0.02$. As noted by \citet{HO10.1}, the differential and normalized cumulative skeleton lengths are equivalent quantities, and should return comparable information about the field.

In order to visually compare the results for the fiducial Gaussian and non-Gaussian cosmologies it is customary to consider the distribution of the difference between the differential skeleton length of light cone realizations in the non-Gaussian models to the average value of the differential skeleton length in the Gaussian model. In other words, for each non-Gaussian cosmology the distribution of the quantity $\mathcal{D}(\eta_\mathrm{t}) - \left\langle \mathcal{D}^\mathrm{(G)}(\eta_\mathrm{t})\right\rangle$ is considered. A similar approach is followed for the normalized cumulative skeleton length, with the trivial replacement of $\mathcal{D}(\eta_\mathrm{t})$ with $\mathcal{C}(\eta_\mathrm{t})$. This is done because the differences between cosmologies that we would like to highlight are usually very small, such that they would be barely be visible in plots such as those shown in Figure \ref{fig:differentialLengthDistribution}. 

\begin{figure}
\begin{center}
	\includegraphics[width=\hsize]{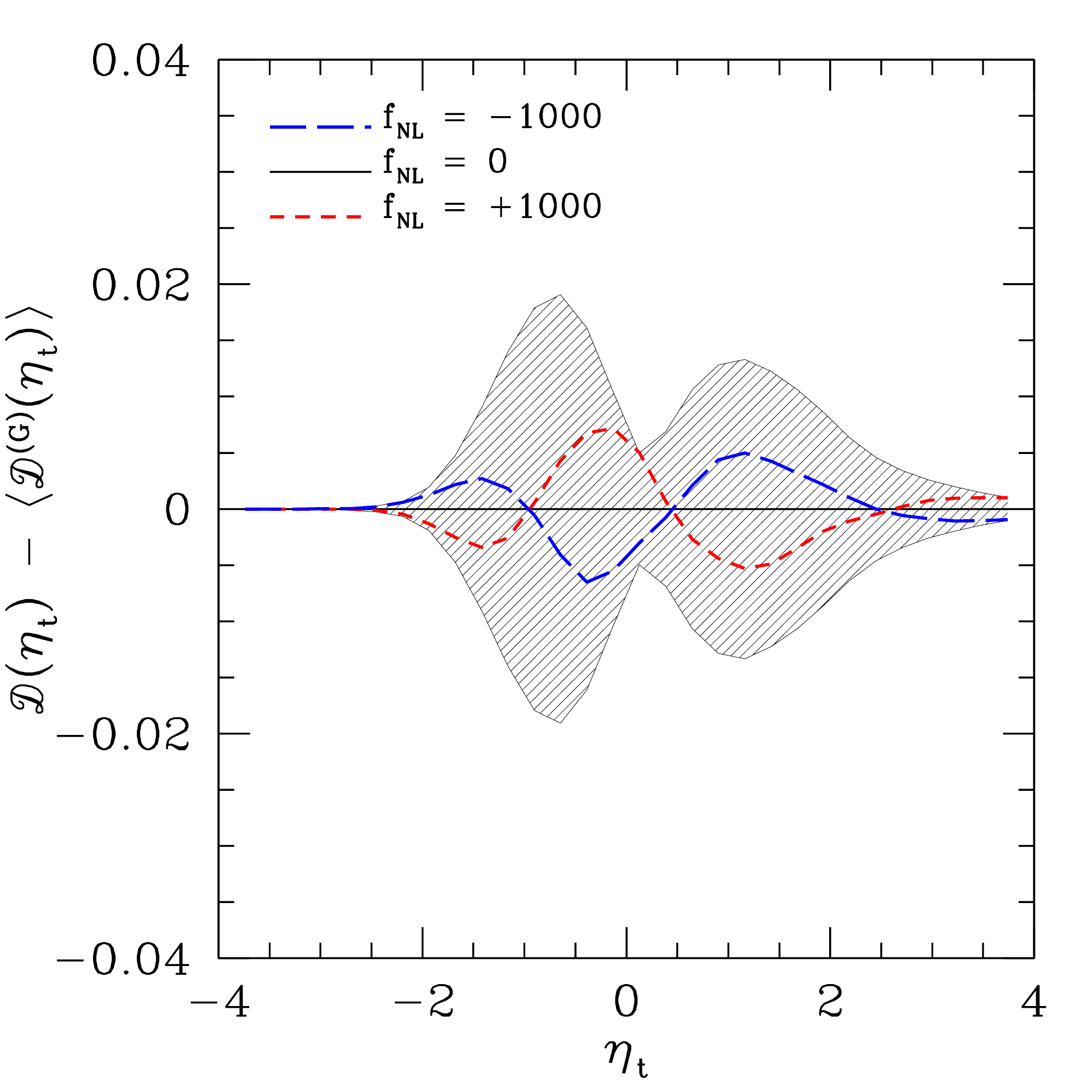}\hfill
	\caption{The difference between the differential skeleton length in the two non-Gaussian cosmologies with $f_\mathrm{NL} = \pm 1000$ and the average value of the same quantity in the reference Gaussian cosmology. The black solid line represents the average value of the distribution for the Gaussian cosmology (that is correctly zero), and the shaded region indicates the dispersion around the average. The other lines refer to the averages for the two non-Gaussian cosmologies, as labeled in the plot. Maps are smoothed with a Gaussian of radius $\ell = 0.002$ times their side length.}
\label{fig:differentialDifferenceDistribution}
\end{center}
\end{figure}

In Figure \ref{fig:differentialDifferenceDistribution} we report the distribution of the difference between the differential skeleton length for the non-Gaussian cosmologies with $f_\mathrm{NL} = \pm 1000$ and the average value of the same quantity in the fiducial Gaussian model. We chose the two most extreme non-Gaussian models in order to better highlight the effects. Also, we refer to the smoothing length $\ell = 0.002$ times the map side length. As can be seen, the differential skeleton lengths in the two non-Gaussian simulations differ on average from their Gaussian counterpart of at most $\sim 0.01$, while the scatter around the mean in the Gaussian case can reach $\sim 0.02$. In other words, the effect of even a large level of primordial non-Gaussianity on the differential skeleton length would be substantially smaller than the field-to-field variance of the fiducial model.

\begin{figure}
\begin{center}
	\includegraphics[width=\hsize]{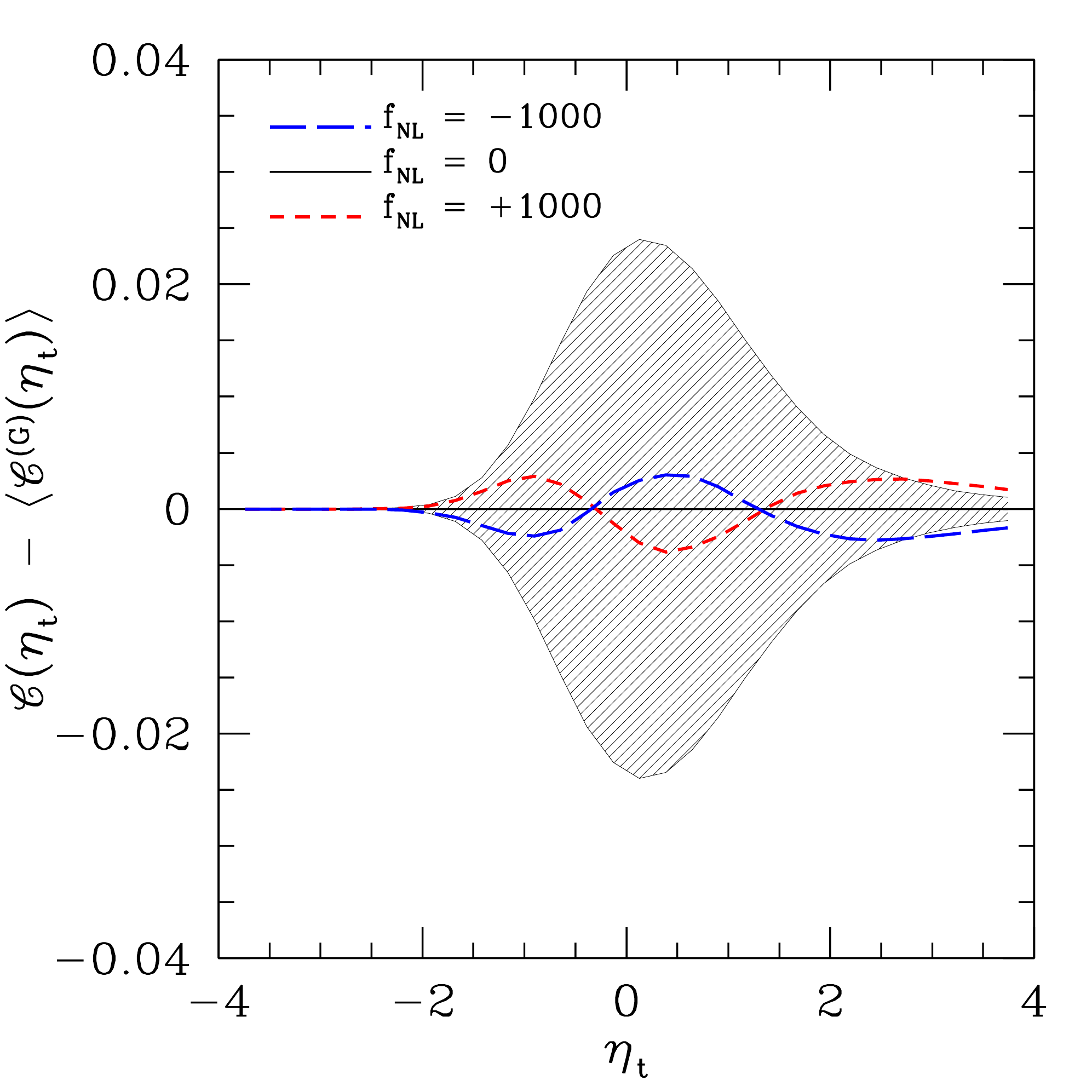}\hfill
	\caption{The difference between the normalized cumulative skeleton length in the two non-Gaussian cosmologies with $f_\mathrm{NL} = \pm 1000$ and the average value of the same quantity in the reference Gaussian cosmology. The black solid line represents the average value of the distribution for the Gaussian cosmology (which is correctly zero), and the shaded region indicates the dispersion around the mean. The other lines refer to the averages of the non-Gaussian cosmologies, as labeled in the plot.}
\label{fig:cumulativeDifferenceDistribution}
\end{center}
\end{figure}

A similar conclusion can be reached by looking at Figure \ref{fig:cumulativeDifferenceDistribution}, where we show the same as in Figure \ref{fig:differentialDifferenceDistribution} but for the cumulative length distribution. Again the effect of primordial non-Gaussianity tends to be significantly smaller than the cosmic variance estimated for the fiducial Gaussian cosmology. In both cases, we decided not to show the results for the larger smoothing length $\ell = 0.02$. These results show the same generic features than those for $\ell = 0.002$, however the cosmic variance is much more important (see the discussion above). Hence we decided to stick just with the most promising case, to which we limit our subsequent analysis.

In Figure 1 of \citet{HO10.1} the authors show the equivalent of our Figures \ref{fig:differentialDifferenceDistribution} and \ref{fig:cumulativeDifferenceDistribution} for a perfectly Gaussian two-dimensional random field that has been perturbed by a non-Gaussian contribution with local shape. There are three main differences between their plots and ours: $i)$ the curves with positive and negative $f_\mathrm{NL}$ are inverted, due to a sign difference in the definition of the differential and cumulative skeleton lengths (irrelevant for our purposes); $ii)$ our effective convergence maps cannot be a non-Gaussian random field with local shape, due to the non-linear growth of structures and the projection procedure; $iii)$ \citet{HO10.1} do not make any distinction between the skeleton and the anti-skeleton. Despite these differences, it is remarkable that their plots look rather similar to ours. Specifically, the succession of minima and maxima of the non-Gaussian curves and the generic shape of the dispersion around the mean for the Gaussian case are well captured. The main difference between the plots of \citet{HO10.1} and ours resides in the fact that ours are not symmetric around $\eta_\mathrm{t} = 0$, and are more skewed toward higher values of the excursion threshold. This is a consequence of the combination of points $ii)$ and $iii)$ above, in that the absence of the anti-skeleton in our analysis gives more importance to the ridges of the effective convergence, while the non-linear clustering of dark-matter enhances the ridges themselves. Despite these differences, we can conclude that the skeleton of the effective convergence keeps a good memory of the statistics of initial density fluctuations, and for this reason it is a potentially powerful mean for probing primordial non-Gaussianity.

\subsection{Statistical analysis}

As a final step of our analysis, we performed a statistical investigation in order to understand what is the level of local primordial non-Gaussianity that could be detected by measuring the skeleton length of cosmic shear maps similar to those simulated here. The analysis is the same for both the differential length and the normalized cumulative length with just the replacement of $\mathcal{D}(\eta_\mathrm{t})$ with $\mathcal{C}(\eta_\mathrm{t})$. Therefore, in the following we describe the formalism in the former case only. 

As a first step, we recall that the differential and normalized cumulative skeleton lengths have been computed by adopting $\nu=32$ different threshold values $\eta_{\mathrm{t},i}$, evenly spaced between $-4$ and $+4$ (we remind the reader that the threshold $\eta_\mathrm{t}$ is in units of the \emph{rms} of $\kappa_\mathrm{e}$). In the following discussion we adopt the notation $\mathcal{D}_i(f_\mathrm{NL})\equiv \mathcal{D}(f_\mathrm{NL},\eta_{\mathrm{t},i})$, and $\mathcal{D}_i^\mathrm{(G)}\equiv \mathcal{D}_i(0)$. Given this, we first of all computed the covariance between different thresholds for the Gaussian model, as

\begin{equation}
\Psi_{ij} = \left\langle \mathcal{D}_i^\mathrm{(G)}\mathcal{D}_j^\mathrm{(G)}  \right\rangle - \left\langle \mathcal{D}_i^\mathrm{(G)}\right\rangle \left\langle\mathcal{D}_j^\mathrm{(G)}  \right\rangle,
\end{equation}
where the average values are taken over the $60$ light cone realizations.

Then, we assumed that the differential skeleton length of the fiducial Gaussian model was the measured one, with each light-cone realization representing an independent measurement, while the average value of the skeleton length for the various non-Gaussian cosmologies was the prediction of the assumed model. We thus defined a $\chi^2(f_\mathrm{NL})$ function according to  

\begin{equation}\label{eqn:chi2}
\chi^2(f_\mathrm{NL})= \sum_{ij=1}^\nu\left[ \left\langle\mathcal{D}_i^{\phantom{i}{}}(f_\mathrm{NL})\right\rangle - \mathcal{D}_i^\mathrm{(G)} \right] \left(\Psi^{-1}\right)_{ij}\left[ \left\langle\mathcal{D}_j(f_\mathrm{NL})\right\rangle - \mathcal{D}_j^\mathrm{(G)} \right].
\end{equation}
In principle, for a fixed value of $f_\mathrm{NL}$, one could compute $\chi^2(f_\mathrm{NL})$ for each of the $60$ light-cone realizations, and then consider the average value. However, we verified that $60$ realizations are not sufficient, in the sense that the average value of $\chi^2(0)$ does not converge to $\nu$ as it is expected to. In order to compensate this, we adopted the following procedure. For each pair $(i,j)$ in the sum of Eq. (\ref{eqn:chi2}) we utilized the values of $\mathcal{D}_i^\mathrm{(G)}$ and $\mathcal{D}_j^\mathrm{(G)}$ of a randomly chosen light-cone. We then repeated this procedure $10^6$ times (obviously allowing for repetitions) and considered the resulting average value of $\chi^2(f_\mathrm{NL})$. Now, we correctly get $\chi^2(0)\simeq \nu=32$.

\begin{figure}
\begin{center}
	\includegraphics[width=\hsize]{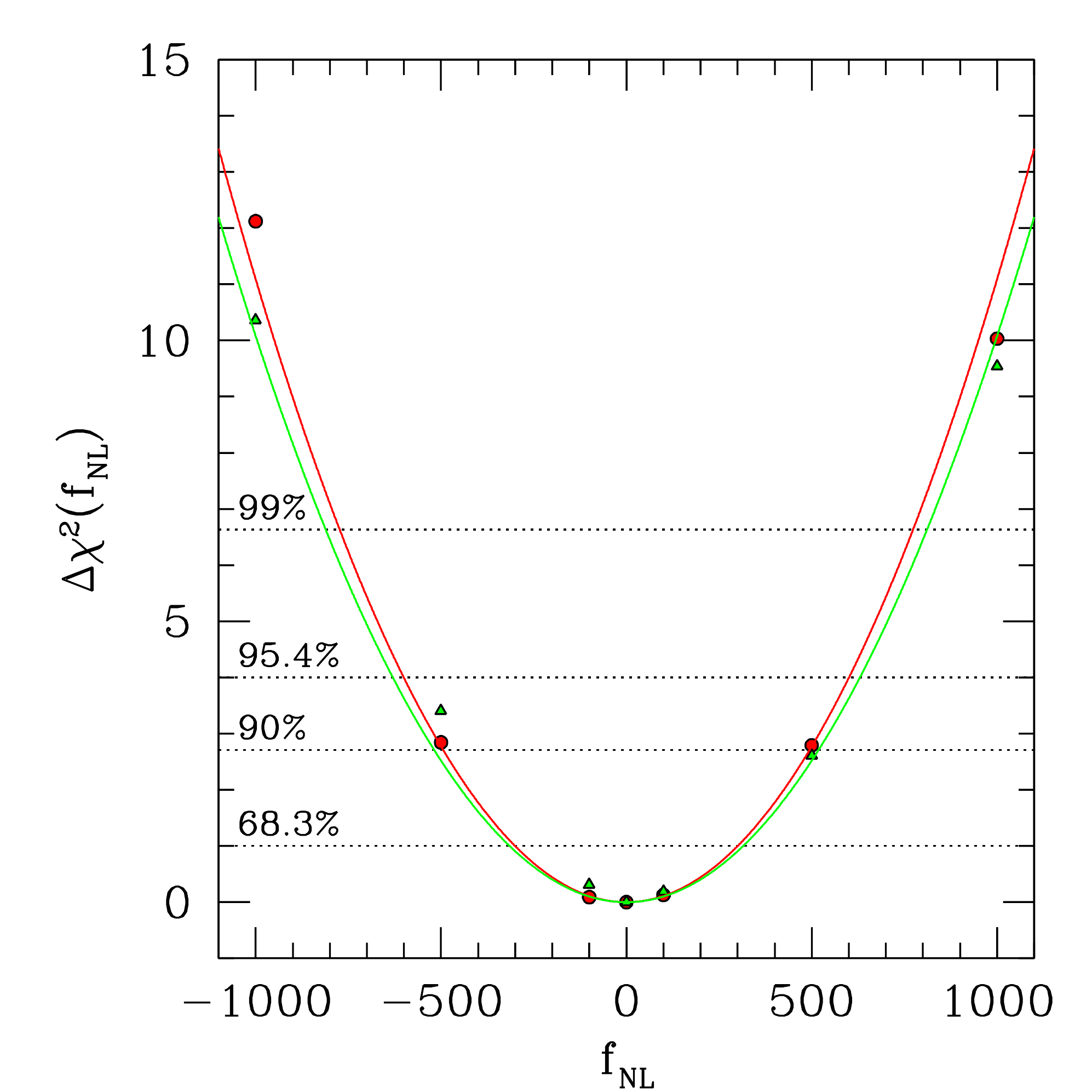}\hfill
	\caption{The $\Delta\chi^2(f_\mathrm{NL})$ function for the differential skeleton length (red line and circles) and the normalized cumulative skeleton length (green line and triangles), assuming a Gaussian smoothing of effective convergence maps with radius $\ell = 0.002$ times their side lengths.
The points represent the actual $\Delta \chi^2$ calculations, while the curves are the corresponding best  parabolic fits. The horizontal dotted lines represent the $\Delta\chi^2$ values corresponding to various Confidence Levels, as labeled.}
\label{fig:chi2}
\end{center}
\end{figure}

We are therefore in the position to compute a $\Delta \chi^2(f_\mathrm{NL})$ function as $\Delta \chi^2(f_\mathrm{NL})=\chi^2(f_\mathrm{NL})-\nu$, that vanishes for $f_\mathrm{NL}=0$, and can thus be used to set confidence intervals on the parameter $f_\mathrm{NL}$. In Figure \ref{fig:chi2} we show the $\Delta\chi^2(f_\mathrm{NL})$ values computed for each of the simulations in our sample, and considering both the differential and the normalized cumulative skeleton lengths. In order to set Confidence Levels (CL henceforth), we fitted the results with two parabolic curves. The fact that the $\Delta\chi^2(f_\mathrm{NL})$ curves are well fitted by parabolae implies the interesting fact that both the differential and the cumulative skeleton lengths are linear functions of $f_\mathrm{NL}$. Moreover, the curves resulting from the differential and the cumulative skeleton lengths are very similar, coherently with the fact that the two estimators are equivalent and should return consistent results. Using standard $\Delta\chi^2$ variations corresponding to one parameter, it turns out that by computing the local skeleton of effective convergence maps covering $\sim 35$ sq. deg would measure the level of primordial non-Gaussianity with an error $\Delta f_\mathrm{NL} \sim 300$ at $68.3\%$ CL, and $\Delta f_\mathrm{NL}\sim 500$ at $90\%$ CL. As explained above, constraints obtained by maps with a smoothing length of $\ell = 0.02$ would be considerably looser, due to the much larger impact of the cosmic variance.

\section{Discussion and conclusions}\label{sct:conclusions}

In this work we computed the local approximation of the skeleton for simulated effective convergence maps. The maps are extracted from cosmological simulations of structure formation with non-Gaussian initial conditions of the local shape. For each non-Gaussian cosmology we have $60$ maps, that can be considered as independent realization of the same light cone, and each map covers an area of $\sim 35$ sq. deg. The simulations span a wide range in the level of primordial non-Gaussianity $f_\mathrm{NL}$, having in particular $f_\mathrm{NL}=0$, $f_\mathrm{NL}=\pm 100$, $f_\mathrm{NL}=\pm 500$, and $f_\mathrm{NL}=\pm 1000$. In order to insure differentiability of the field we smoothed the maps with a Gaussian having two different length scales, namely $\ell = 0.02$ and $\ell = 0.002$ times the side length of the maps themselves.

By considering excursion subsets of each map we evaluated the differential and cumulative length of the local skeleton, which carry equivalent cosmological information, yet alternative to more common statistics like the PDF or the peak number counts of the field. The main results that we derived in this paper can be summarized as follows.

\begin{itemize}
\item The total skeleton length of effective convergence maps is larger in models with a positive $f_\mathrm{NL}$ and smaller if $f_\mathrm{NL}$ is negative. The separation between models with positive and negative $f_\mathrm{NL}$ values tends to vanish if the field is smoothed with a relatively large smoothing length. This is so because in this case the most prominent ridges of the field, that are presumably the most affected by primordial non-Gaussianity, tend to be washed out.
\item The result outlined in the previous point is in agreement with the fact that a positive-skewed primordial non-Gaussianity leads to an anticipated formation of structures, with the filamentary structure being thus more pronounced. The subsequent flow of matter from the filaments to the knots of the cosmic web, which should also be enhanced in models with positive $f_\mathrm{NL}$, is not enough to counteract the first effect, at least at the redshifts probed by cosmological weak lensing.
\item The distributions of the difference between the normalized cumulative and differential skeleton lengths and the average values of the same quantities in the fiducial Gaussian cosmology are qualitatively rather similar to what one would expect for a Gaussian random field perturbed according to a local shape non-Gaussianity.  This implies that these statistics retain good memory of the initial conditions, despite the subsequent non-linear growth of structures and the projection procedure.
\item Even if the most extreme non-Gaussian models are used, having $f_\mathrm{NL} = \pm 1000$, the average differences of the differential and normalized cumulative skeleton lengths to the corresponding average values in the Gaussian case are always significantly smaller than the dispersion around the mean for Gaussian cosmologies. In other words, the signal we are trying to detect is always smaller than the noise given by the cosmic variance for the field size we have considered. This is more so for larger smoothing scales, since smoother maps are expectedly more affected by field-to-field variations.
\item A simple statistical investigation based on $\chi^2$ intervals shows that the constraints expected by the skeleton analysis performed on cosmic shear maps similar to those used here are rather weak. Particularly, the expected error on the level of primordial non-Gaussianity is of $\Delta f_\mathrm{NL} \sim 300$ at $68.3\%$ CL and of only $\Delta f_\mathrm{NL} \sim 500$ at $90\%$ CL. These numbers are substantially larger than the current constraints given by the CMB and LSS.
\end{itemize}

While the results concerning the constraints on $f_\mathrm{NL}$ can be quite discouraging, it is worth stressing that they are valid only for the map size that has been considered here, $\sim 35$ sq. deg. Evidently, larger maps would be less affected by cosmic variance, and thus would lead to more stringent constraints on the level of primordial non-Gaussianity. For the same reason, it is safe to state that measuring the skeleton of the effective convergence in the COSMOS field ($\sim 2$ sq. deg) or in the \emph{Deep Lens} field ($\sim 4$ sq. deg) would not lead to any signature of primordial non-Gaussianity. Although we are not in a position to properly quantify how much the covered area should be in order to get significant bounds on $f_\mathrm{NL}$, we can perform a rough estimate along the following lines. In \citet{SO04.1} and \citet{MO11.1} the authors quantify how the cosmic variance for dark matter density fluctuations scales with the volume of space that is probed. It turns out that $\sigma^2_\mathrm{c} \propto V^{-\gamma}$ with $\gamma \sim 0.4-0.6$, although a power law is probably not the best description of the trend. For a fixed source redshift, the weak lensing-surveyed volume of space is proportional to the covered area, thus we can infer that the same scaling applies to the sky coverage as well. Now, we are not looking at the dark matter density directly, rather at the skeleton of its projection. Still, given the similarities between, e.g., the differential skeleton length and the PDF of the effective convergence, we assume that the relative cosmic variance shares the same scaling with the survey area. This means that, for instance, the CFHTLS field, covering almost $5$ times the area that we considered here ($\sim 170$ sq. deg), would lead to $\Delta f_\mathrm{NL} \sim 200$ at $68.3\%$ CL. In order to have an error on the estimated value of primordial non-Gaussianity of $\Delta f_\mathrm{NL} \sim 100$ at the same CL, we would need a sky coverage of $\sim 3000$ sq. deg. Planned wide field cosmic shear surveys such as \emph{Euclid} \citep{LA09.1} will produce effective convergence maps of the entire extragalactic sky, thus providing a powerful test of the initial conditions by using the skeleton statistics, with an estimated $\Delta f_\mathrm{NL} \sim $ a few tens at $68.3\%$ CL.

Meanwhile, we would also like to outline a few limitations of our approach. We assumed perfect knowledge of the effective convergence in our maps, while real maps are affected by various sources of noise, both systematic and statistical, such as the intrinsic ellipticity distribution of the sources used to measure the cosmic shear. Furthermore, in our statistical analysis we were allowed to vary only the parameter $f_\mathrm{NL}$, while a more complete investigation would consider also marginalizing over the other cosmological parameters (especially $\sigma_8$) that might affect the skeleton length. Performing this, and properly answering the above question about the minimum field size necessary to obtain competitive constraints on $f_\mathrm{NL}$ would require a much larger and detailed set of simulations, that is not available at the moment.

As this work was being finalized, another paper dealing with the topology of weak-lensing maps appeared on the archive, \citet{MU11.1}. That work is a nice complement to our, since it presents a perturbative approach in order to compute the Minkowski Functionals of the effective convergence. It further stresses the importance of investigating all pieces of information extractable from cosmological weak lensing (morphology, correlation function, peak counts, etc.), in order to maximize the return of present and future cosmic shear surveys.

\section*{Acknowledgements}
CF is grateful to A.H. Gonzalez for useful discussions on the manuscript. LM acknowledges financial contributions from contracts ASI-INAF I/023/05/0, ASI-INAF I/088/06/0, ASI I/016/07/0 'COFIS', ASI 'Euclid-DUNE' I/064/08/0, and PRIN MIUR 'Dark energy and cosmology with large galaxy surveys'. KD acknowledges the support by the DFG Priority Programme 1177 and additional support by the DFG Cluster of Excellence 'Origin and Structure of the Universe'. This work was sustained in part by the German Academic Exchange Service (DAAD) under the Vigoni programme. We are grateful to an anonymous referee for comments that helped us improving the presentation of our work.

{\small
\bibliographystyle{aa}
\bibliography{./master}
}

\end{document}